\documentclass[aps,prd,preprint,groupedaddress,showpacs,nofootinbib]{revtex4}

\usepackage{amsmath}
\usepackage{amssymb}
\usepackage{mathtools}
\usepackage{graphicx}

\newcommand{\beq}{\begin{equation}}
\newcommand{\eeq}{\end{equation}}
\newcommand{\bea}{\begin{eqnarray}}
\newcommand{\eea}{\end{eqnarray}}

\begin{document}

\title{Inflation from cosmological constant and
 nonminimally coupled scalar}

\author{Dra\v zen~Glavan$^*$, Anja Marunovi\'c$^*$ and Tomislav Prokopec}
\email[]{E-mail: d.glavan@uu.nl; a.marunovic@uu.nl;
t.prokopec@uu.nl}

\affiliation{Institute for Theoretical Physics, Spinoza Institute
and
 Center for Extreme Matter and Emergent Phenomena,
 Utrecht University, Postbus 80.195,
  3508 TD Utrecht, The Netherlands}


\begin{abstract}
 We consider inflation in a universe with a positive cosmological constant and a nonminimally coupled
scalar field, in which the field couples both quadratically and
quartically to the Ricci scalar. When considered in the Einstein
frame and when the nonminimal couplings are negative, the field
starts in slow roll and inflation ends with an asymptotic value of
the principal slow roll parameter, $\epsilon_E=4/3$. Graceful exit
can be achieved by suitably (tightly) coupling the scalar field to
matter, such that at late time the total energy density reaches the
scaling of matter, $\epsilon_E=\epsilon_m$. Quite generically the
model produces a red spectrum of scalar cosmological perturbations
and a small amount of gravitational radiation. With a suitable
choice of the nonminimal couplings, the spectral slope can be as
large as $n_s\simeq 0.955$, which is about one standard deviation
away from the central value measured by the Planck satellite. The
model can be ruled out by future measurements if any of the
following is observed: (a) the spectral index of scalar
perturbations is $n_s>0.960$; (b) the amplitude of tensor
perturbations is above about $r\sim 10^{-2}$; (c) the running of the
spectral index of scalar perturbations is positive.

\end{abstract}


\maketitle


\section{Introduction}
\label{sec:Introduction}

 The most famous example of an inflationary model realized within
a tensor-scalar (TeS) theory~\cite{Jordan:1955,Brans:1961sx,Bergmann:1968ve,Fakir:1990eg},
in which a (gravitational) scalar couples to the Ricci scalar, is
Higgs inflation~\cite{Salopek:1988qh,Bezrukov:2007ep,Bezrukov:2013fka,Prokopec:2014iya},
in which the role of the inflaton is played by the standard model Higgs field.
Tensor-scalar theories have also been extensively used to discuss the cosmological constant
problem~\cite{Dolgov:1982gh,Ford:1987de,Weinberg:1988cp,Glavan:2015ora}
to explain the origin of dark energy~\cite{Glavan:2014uga,Glavan:2015,Copeland:2006wr,Peebles:2002gy}
 and have been thoroughly tested on solar system scales~\cite{Will:2005va}.

  While many inflationary models have been considered, to our knowledge no one has investigated the model
in which inflation is driven by a positive cosmological constant
accompanied by a nonminimally coupled scalar field. A study of this
class of models is the subject of this paper.

In section~\ref{sec:The model} we present the model and
discuss how to analyze it in the Einstein frame.  In section~\ref{sec:Slow roll inflation}
we recall the basics of slow roll approximation. In section~\ref{sec:Results}
our principal results are presented. In particular,
we discuss  the spectral index, its running and the amplitude of tensor perturbations.
Finally, in section~\ref{sec:Discussion} we shortly recapitulate our main results and discuss future directions.
A particular emphasis is devoted to the graceful exit problem and to  the question of falsifiability of our inflationary model.


\section{The Model}
\label{sec:The model}

 In this paper we consider the following simple tensor-scalar theory of gravity,
 whose action in the Jordan frame reads,
\begin{equation}
  S_J = \int d^4 x\sqrt{-g_J}\left(\frac12 F(\phi_J)R_J - M_{\rm P}^2\Lambda
   - \frac12 g_J^{\mu\nu}\partial_\mu\phi_J\partial_\nu\phi_J-V_J(\phi_J)\right)
\,,
\label{action:J}
\end{equation}
where $g_J ={\rm det}[g_{J\mu\nu}]$,  $g_J^{\mu\nu}$ is the inverse
of the (Jordan frame) metric tensor $ g_{J\mu\nu}$ and $R_J$ is the
Ricci scalar. In this paper we assume the following simple form for
the function $F$ and $V_J$,
\begin{equation}
   F(\phi_J) = M_{\rm P}^2 - \xi_2\phi_J^2-\xi_4\frac{\phi_J^4}{M_{\rm P}^2}
\,,\qquad  {\rm and}\quad  V_J(\phi_J)=0
\,,
\label{F and V}
\end{equation}
where $M_{\rm P}^2=1/(8\pi G_N)$, $G_N$ is the Newton constant and
$\xi_2$ and $\xi_4$ are (dimensionless) nonminimal coupling parameters.
In our conventions conformal coupling corresponds to $\xi_2\equiv \xi_c=1/6$, $\xi_4=0$,
 and we work with natural units in which $\hbar = 1 = c$. For the metric
we take a cosmological, spatially flat, background,
\begin{equation}
 g_{J\mu\nu} = {\rm diag}[-1,a_J^2(t),a_J^2(t), a_J^2(t)]
\,.
\label{metric}
\end{equation}

Even though the Jordan and Einstein frames are fully
equivalent~\cite{Prokopec:2014iya,Weenink:2010rr,Prokopec:2013zya,Prokopec:2012ug},
cosmological perturbations are easier to analyze in the Einstein frame and when a slow roll approximation is
utilized. Therefore, we shall proceed by transforming the Jordan frame action~(\ref{action:J})
to the Einstein frame.

 To get to the Einstein frame with the canonically coupled scalar, one ought to perform the following frame (conformal) transformations,
 \begin{equation}
  g_{E\mu\nu} = \frac{F(\phi_J)}{M_{\rm P}^2}g_{J\mu\nu}
  \,,\qquad d\phi_E = \frac{M_{\rm P}}{F(\phi_J)}
  \sqrt{F(\phi_J)+\frac32\bigg(\frac{dF(\phi_J)}{d\phi_J}\bigg)^{\!2}}
 \;d\phi_J
 \,,
 \label{conformal frame transformations}
 \end{equation}
where the index $E$ refers to the Einstein frame. In this frame, the
scalar-tensor action~(\ref{action:J}) becomes
simpler~\cite{Glavan:2015ora},
\begin{equation}
  S_E = \int d^4 x\sqrt{-g_E}\left(\frac{M_{\rm P}^2}{2}R_E - \frac12 g_E^{\mu\nu}\partial_\mu\phi_E\partial_\nu\phi_E
- \frac{M_{\rm P}^6\Lambda}{F^2(\phi_J(\phi_E))}\right)
\,,
\label{action:E}
\end{equation}
thus coupling the cosmological constant to the scalar field. This
coupling introduces a nontrivial dynamics which -- as we show below
-- can be used to realize a viable model of primordial inflation.
\begin{figure}[th]
\begin{minipage}[t]{.45\textwidth}
        \begin{center}
\includegraphics[scale=0.6]{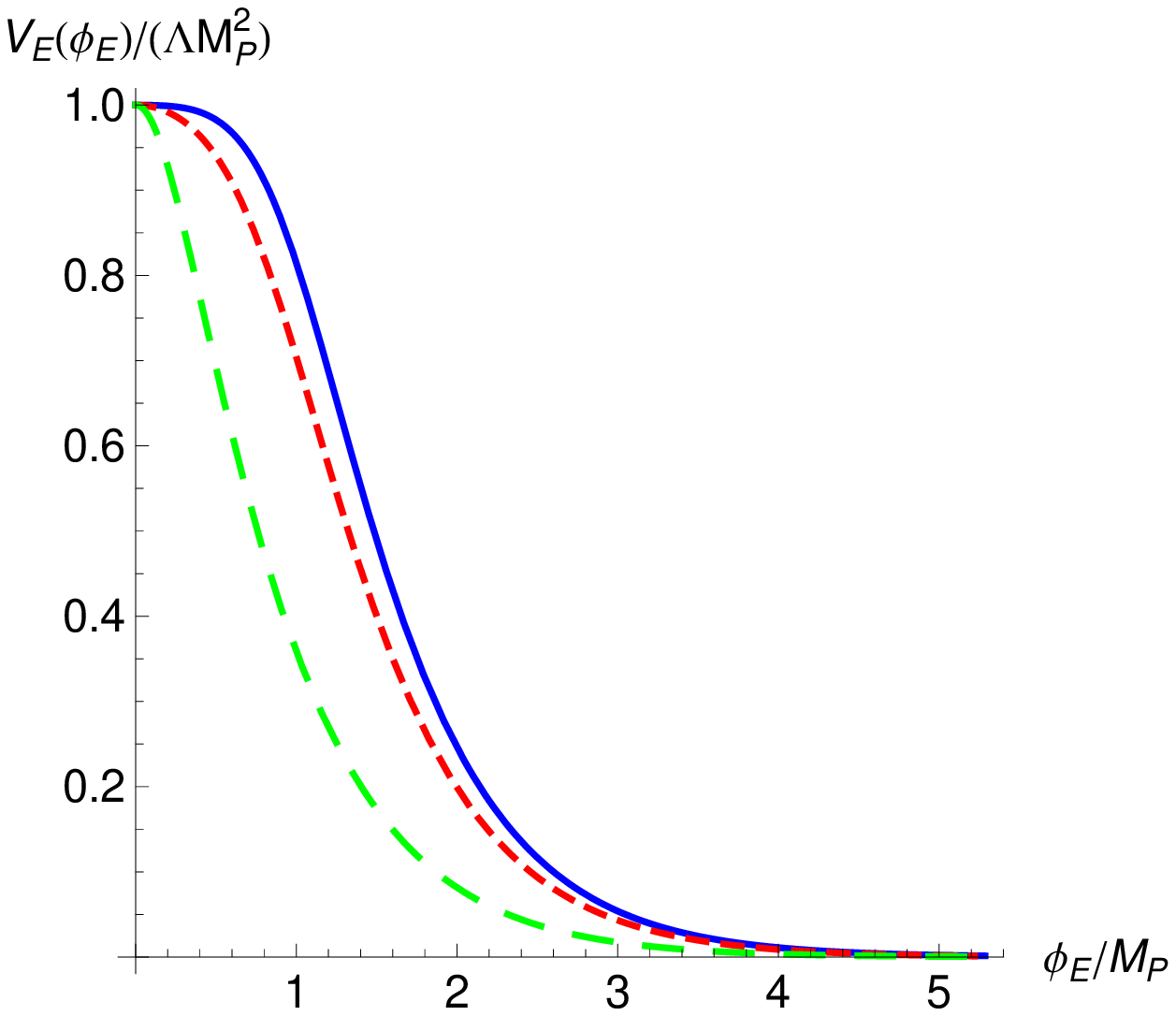}
\end{center}m
    \end{minipage}
\begin{minipage}[t]{.45\textwidth}
        \begin{center}
\includegraphics[scale=0.6]{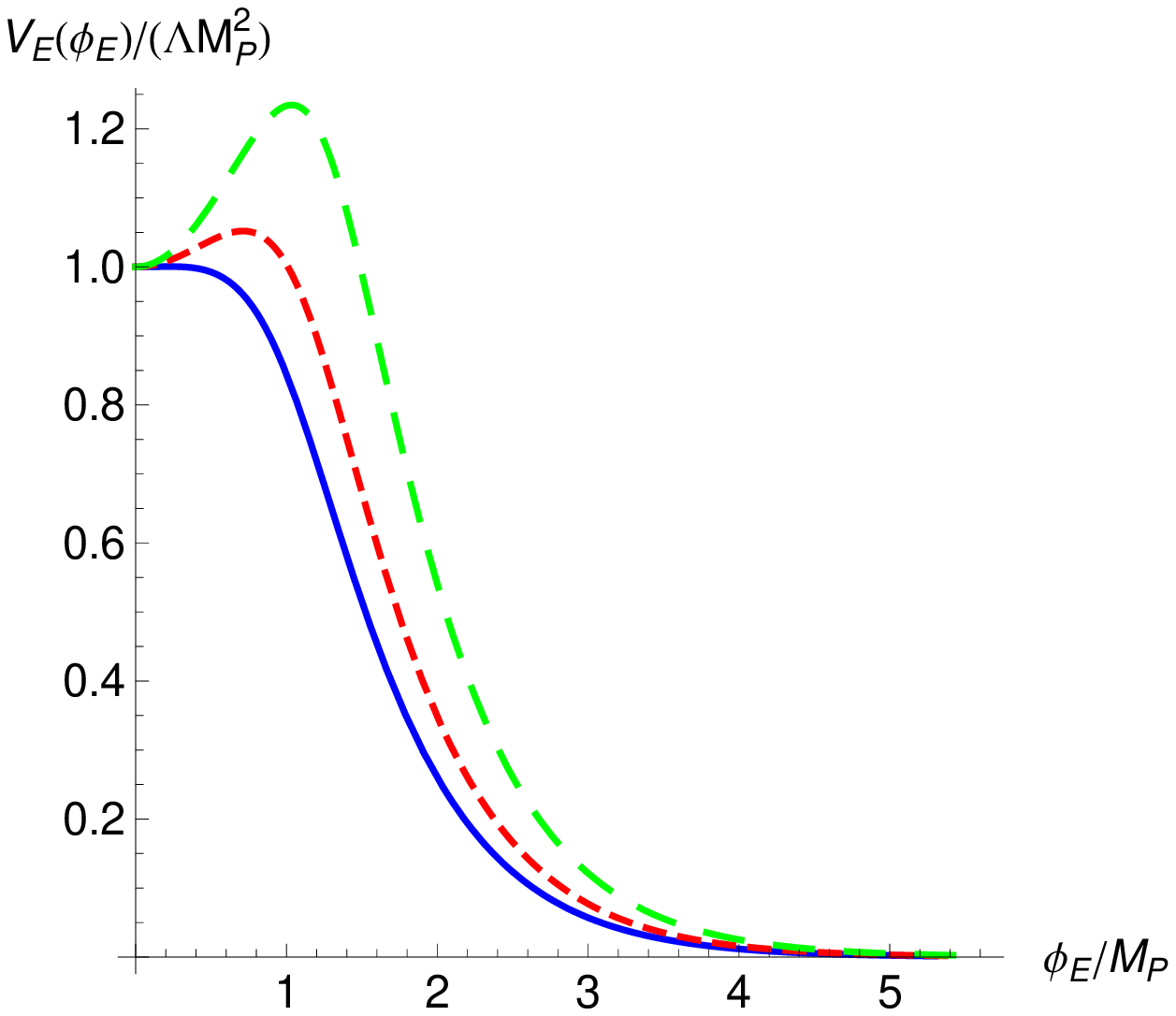}
\end{center}
    \end{minipage}
\caption{The effective potential (cosmological constant) $V_E$ in the Einstein frame as  a
function of the Einstein frame field $\phi_E$. In this figure $\xi_4=-0.1$.
{\it Left panel:}    $\xi_2=-0.01$ (blue solid), $\xi_2=-0.1$  (red dashes) and $\xi_2=-1$  (long green dashes).
{\it Right panel:}   $\xi_2=0.01$ (blue solid), $\xi_2=0.1$  (red dashes) and $\xi_2=0.2$  (long green dashes).
Note that when $\xi_2<0$,  $V_E$ has a local maximum at $\phi_E=0$ ($\phi=0$), while
for $\xi_2>0$, $V_E$ has a local minimum at $\phi_E=0$ and two local maxima at  some $\phi_E=\pm\phi_{E0}\neq 0$.
The potential $V_E$ exhibits a $Z_2$ symmetry, {\it i.e.} it is symmetric under $\phi_E\rightarrow -\phi_E$.}
\label{VE as function of phiE and xi2}
\end{figure}

In figure~\ref{VE as function of phiE and xi2} we show the effective potential in the Einstein frame
 $V_E(\phi_E)=M_{\rm P}^6\Lambda/F^2(\phi_J(\phi_E))$ as a function of
the Einstein frame field $\phi_E$ for several values of $\xi_2$ and
for $\xi_4$ fixed to $\xi_4=-0.1$. When both couplings are negative,
the effective potential has one local maximum (at $\phi_E=0$) and it
decays monotonically towards zero as the field $|\phi_E|$ increases
(see left panel). However, when $\xi_2>0$ and $\xi_4<0$, $V_E$
develops a local minimum at $\phi_E=0$ and two local maxima at some
positive  $|\phi_E|$ (right panel). In this paper we investigate the
case when both couplings are negative and leave the latter case, in
which tunneling from the local minimum can play an important role,
for future work. While the field dependence of the potential
in~(\ref{action:E}) is simple when expressed in terms of the Jordan
frame field, there is no simple analytic form that describes the
Einstein frame potential. This is a consequence of the fact
that~(\ref{conformal frame transformations}) cannot be solved
analytically for $\phi_J(\phi_E)$.  There are simple limits however.
For small field values, $\phi_E\ll M_{\rm P}$, the potential $V_E$
in~(\ref{action:E}) can be approximated by a constant plus a
negative mass term (as in hilltop inflation, see {\it e.g.}
Refs.~\cite{Boubekeur:2005zm,Lyth:1998xn}),
\begin{equation}
V_E(\phi_E)\simeq \Lambda\left[M_{\rm P}^2+2\xi_2\phi_E^2\right] +{\cal O}(\phi_E^4)
   \,,
\label{VE small field}
\end{equation}
while for $\phi_E\gg M_{\rm P}$, the potential decays exponentially with the field,
\begin{equation}
V_E(\phi_E)\simeq V_{E0}\exp\left(\!-\!\lambda_E\frac{\phi_E}{M_{\rm P}}\right)
\,,\quad
 \lambda_E = \sqrt{\frac{8}{3}}
  \,,
\label{VE large field}
\end{equation}
where $V_{E0}$ is a constant whose value is $\sim\Lambda M_{\rm
P}^2$. From Eqs.~(\ref{VE small field}) and~(\ref{VE large field})
we see that, if the field starts from some small value near the
local maximum, it will slowly roll down the hill, exiting eventually
inflation when $\epsilon_E\simeq 1$.

One can show~\cite{Glavan:2015ora} that for $\xi_4=0$, $\xi_2<-1/2$
and in the Einstein frame
\beq \epsilon_E=\frac{-8\xi_2}{1-6\xi_2}>1\,, \label{epsilonInfty}
\eeq
with the limiting value $\epsilon_E\rightarrow 4/3$ for
$\xi_2\rightarrow-\infty$. Here we have introduced quartic
nonminimal coupling $\xi_4<0$ in $F$ in Eq.~(\ref{F and V}) in order
to be able to relax the condition on $\xi_2$ and to still be able to
terminate inflation. Namely, one can show that even when the quartic
coupling is arbitrarily small and negative, $\epsilon_E$ will
asymptotically reach the value $4/3>1$, regardless of the value of
negative $\xi_2$. The condition $\epsilon_E\ll 1$ during inflation
requires $|\xi_2|\ll 1$ which is satisfied by this setup.

\begin{figure}[hht]
        \begin{center}
\includegraphics[scale=0.7]{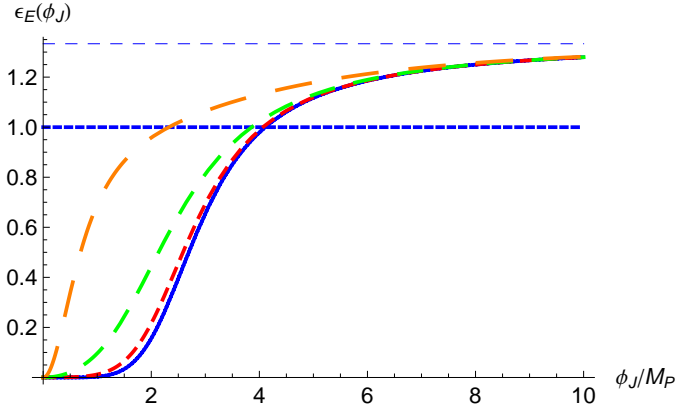}
\vskip -2cm
\end{center}
\caption{Principal slow roll parameter $\epsilon_E$  as a function
of $\phi_J$ for $\xi_4=-0.01$. Different curves show: $\xi_2=-0.01$
(blue solid), $\xi_2=-0.02$ (short red dashes), $\xi_2=-0.1$ (green
dashes) and  $\xi_2=-0.5$ (long orange dashes). Note that,
independently on the values of $\xi_2$  and $\xi_4$ (as long as they
are both negative), $\epsilon_E\rightarrow 4/3$ (horizontal blue
dashes) when $\phi\rightarrow \infty$. Inflation ends when
$\epsilon_E\rightarrow 1$  (short horizontal blue dashes).}
\label{epsE as function of phi}
\end{figure}

One way of seeing this is to work in the adiabatic approximation and
subsume the $-\xi_4\phi^4/M_{\rm P}^2$ term in $F$ into a field
dependent quadratic coupling $\xi$ as follows: $\xi(\phi)\equiv
\xi_2 + \xi_4\phi^2/M_{\rm P}^2$. Now, when this is inserted into
$\epsilon_E\simeq -8\xi/(1-6\xi)$, which is the attractor value at
asymptotically large field values, one obtains,
$\epsilon_E\rightarrow 4/3$ for arbitrarily small, negative values
of $\xi_4$, see figure~\ref{epsE as function of phi}.

While inflation terminates when $\epsilon_E>1$, $\epsilon_E=4/3$ is
not enough to explain the post-inflationary radiation and matter
eras. One can show~\cite{Glavan:2015ora} that a suitable coupling to
a (perfect) matter fluid can induce the decay of $\phi_E$ into
matter, such that in the tightly coupled regime, the system reaches
$\epsilon_E=\epsilon_m$. When matter is predominantly in the form of
a relativistic fluid, for which the equation of state parameter
$w_m={p}_m/\rho_m=1/3$, or equivalently $\epsilon_m=(3/2)(1+w_m)=2$,
one will eventually reach a postinflationary radiation era,
providing thus a graceful exit from inflation that is consistent
with all observations.


 \section{Slow roll inflation}
 \label{sec:Slow roll inflation}

 We do not know what was the state of the Universe before inflation. It seems reasonable to
assume that the Universe was expanding and that it was in a chaotic
state, whose energy-momentum tensor was dominated by field
fluctuations of various (energy and distance) scales. Even if not in
equilibrium, such a state could be approximated by a nearly perfect
fluid, whose equation of state is well approximated by the radiation
equation of state, $w\simeq 1/3$. In such a state nonminimal
couplings do not play a significant role (since $\langle
R\rangle\sim 0$), and thence it is natural to take the expectation
value of the (quantum) field $\hat \phi$ to be close to zero,
 $\langle\hat\phi(x)\rangle=\phi_0\simeq 0$.

 As the Universe expands, the amplitude of fluctuations
decreases, and the corresponding energy density and pressure decrease accordingly, reaching eventually
the point when the contribution from the cosmological constant (whose origin may be both
geometric and vacuum fluctuations of quantum fields) becomes significant. At that moment the Universe
enters an inflationary phase, whereby the field feels a hilltop-like potential~(\ref{VE small field}) and starts rolling
down the hill. As it rolls, the contribution from fluctuations will rapidly redshift, becoming
less and less important for the Universe's dynamics. Thus we see that in our inflationary model the Universe
enters inflation from a broad range of initial conditions without any need for (fine) tuning. Of course, it is still true
that the cosmological constant and the nonminimal couplings have to have the right values
(set by the COBE normalization of the amplitude of the scalar spectrum of cosmological perturbations and
by the Planck value of the corresponding spectral slope). As we show
below, these values can be obtained in our model by quite a natural choice of the parameters.

The Einstein frame action~(\ref{action:E}) implies the following equations of motion,
\begin{eqnarray}
  &&\hskip -1cm
  \ddot \phi_E + 3H_E\dot\phi_E + V_E^\prime = 0\,,
\label{EOM: Einstein frame:1}\\
  H_E^2 &=& \frac{1}{3M_{\rm
  P}^2}\bigg(\frac{\dot\phi_E^2}{2}+V_E(\phi_E)\bigg)\,,
\label{EOM: Einstein frame:2}\\
\dot H_E &=& -\frac{\dot\phi_E^2}{2M_{\rm P}^2}
\,,
\label{EOM: Einstein frame:3}
\end{eqnarray}
where the metric tensor is now, $g_{E\mu\nu}={\rm diag}[-1,a_E^2(t),a_E^2(t),a_E^2(t)]$.
 While these equations can be solved numerically~\cite{Glavan:2015ora} without resorting to slow roll approximation
(in which the Hubble parameter and possibly some of its time
derivatives can be treated as adiabatic functions of time), it is
instructive to use slow roll approximation because one can use
analytical techniques that allow us to get a better grasp of the
parameter dependencies of the observables. One can check the
predictions of slow roll approximation by studying (approximate or
exact) solutions of the attractor equation,
\begin{equation}
  H_E^2(\phi_E) = \frac23 M_{\rm P}^2\bigg(\frac{dH_E}{d\phi_E}\bigg)^2 +\frac{V_E(\phi_E)}{3M_{\rm P}^2}
\,, \label{Hamilton-Jacobi-type}
\end{equation}
which is more general than slow roll approximation.
 This equation  can be derived as follows. In
general $H_E=H_E(\phi_E,\dot\phi_E)$. However, it is often the case
that the dependence on $\dot\phi_E$ can be neglected because the
initial conditions for $\dot\phi_E$ are forgotten or $\dot\phi_E$ is
a function of $\phi_E$ (as it is, for example, in slow roll). More
generally this will be the case when there is a phase space
attractor towards which trajectories $(\phi_E(t),\dot\phi_E(t))$
rapidly converge.~\footnote{An attractor behavior is opposite from a
chaotic behavior, in which phase space trajectories repulse each
other in the sense that they (exponentially) diverge from each
other.} In this case $H_E=H_E(\phi_E)$ and
Eq.~(\ref{Hamilton-Jacobi-type}) can be easily obtained by
rewriting~(\ref{EOM: Einstein frame:3}) as $\dot\phi_E=-2M_{\rm
P}^2dH_E/d\phi_E$ and inserting it into~(\ref{EOM: Einstein
frame:2}). With these caveats in mind, solving
Eq.~(\ref{Hamilton-Jacobi-type}) is equivalent to solving the full
system of equations~(\ref{EOM: Einstein frame:2}--\ref{EOM: Einstein
frame:3}) (Eq.~(\ref{EOM: Einstein frame:1}) does not provide any
new information as it can be obtained from the other two equations).

 In slow roll approximation one neglects the first term in Eq.~(\ref{EOM: Einstein frame:1})
and the kinetic term in Eq.~(\ref{EOM: Einstein frame:2}) (the last
equation is irrelevant because it is not independent). The memory of
the initial conditions is neglected (because $\dot\phi_E$ and $\dot
H_E$ are not independent variables and slow roll is an attractor).
Moreover, the dependence on the initial field value
$\phi_{J0}=\phi_J(t_0)$ is irrelevant, because one measures the
number of e-folds from the end of inflation $\phi_J(t_e)=\phi_{Je}$
(at which the principal slow roll parameter $\epsilon_E=1$), and
during inflation we are in the attractor. With these in mind, we can
define the number of e-folds as,
\begin{eqnarray}
 N(\phi_J) &=& \int_t^{t_e}H_E(\tilde t\,)d\tilde t
      \simeq \frac 12\int_{\tilde\phi_J}^{\phi_{Je}}d\phi_J\bigg[\frac32\frac{F^\prime}{F}+\frac{1}{F^\prime}\bigg]
\nonumber\\
  &=& \frac34\ln\bigg(\frac{F(\tilde\phi_J)}{M_{\rm P}^2}\bigg)
      +\frac{1}{8\xi_2}\ln\bigg(\frac{M_{\rm P}^2F^\prime(\tilde\phi_J)}{\tilde\phi_J^3}\bigg)
              \left|_{\phi_J}^{\phi_{Je}}\right.
\,,
\label{number e-folds}
\end{eqnarray}
where we made use of,
\begin{equation}
 \frac{\dot \phi_E}{H_E} = \frac{2M_{\rm P}F^\prime(\phi_J)}{\sqrt{F(\phi_J)+\frac32{F^\prime(\phi_J)}^2}}
\,.
\label{dot phi E}
\end{equation}
Next, the principal slow roll parameter, $\epsilon_E\equiv \epsilon_1=-\dot H_E/H_E^2$ reads in slow roll approximation,
\begin{equation}
  \epsilon_E(\phi_J) \simeq \frac{2{F^\prime}^2}{F+\frac32{F^\prime}^2}
\,.
\label{eps E}
\end{equation}
The other two slow roll parameters can be defined in terms of the rate of change of $\epsilon_E$ as,
$\eta_E\equiv \epsilon_2 = \dot \epsilon_E/(\epsilon_E H_E)$,
$\xi_E\equiv \epsilon_3 = \dot \eta_E/(\eta_E H_E)$. In slow roll approximation they read,
\begin{eqnarray}
 \eta_E(\phi_J) & \simeq&
 \frac{2F(2F{F^{\prime\prime}-{F^\prime}^2)}}{\big(F+\frac32{F^\prime}^2\big)^2}\,,
\nonumber\\
 \xi_E(\phi_J) &\simeq& \frac{2F^\prime}{F+\frac32{F^\prime}^2}
\Bigg(\frac{2F^2F^{\prime\prime}}{2FF^{\prime\prime}-{F^\prime}^2}
                    - \frac{F^\prime\big(F-\frac32{F^\prime}^2+6F^{\prime\prime}\big)}{F+\frac32{F^\prime}^2}\Bigg)
\,.
\label{slow roll parameters 2 and 3}
\end{eqnarray}
The scalar and tensor perturbations are of the form,
%
\begin{eqnarray}
\Delta_s^2(k)&=& \Delta_s^2(k_*)\bigg(\frac{k}{k_*}\bigg)^{n_s-1}
\,,\qquad   \Delta_s^2(k_*) = \frac{H_E^2}{8\pi^2\epsilon_EM_{\rm
P}^2}\,,
\nonumber\\
 \Delta_t^2(k)&=& \Delta_t^2(k_*)\bigg(\frac{k}{k_*}\bigg)^{n_t}
\,,\qquad   \Delta_t^2(k_*) = \frac{2H_E^2}{\pi^2M_{\rm P}^2}
\,,
\label{two spectra}
\end{eqnarray}
where (to the leading order in slow roll approximation) the spectral
indices $n_s$ and $n_t$ can be determined from the variation of
$\Delta_s^2(k)$  and  $\Delta_t^2(k)$ with respect to $k$ at the
first horizon crossing during inflation (where $k=k_*=Ha$) as
follows,
\begin{eqnarray}
  n_s &=& 1 + \bigg(\frac{d\ln[\Delta_s^2(k)]}{d\ln(k)}\bigg)_{k=k_*}  = \frac{dt}{d\ln(Ha)}\frac{d\ln[\Delta_s^2(k_*)]}{dt} \simeq -2\epsilon_E-\eta_E
\,,
\label{ns}\\
n_t &=& \bigg( \frac{d\ln[\Delta_t^2(k)]}{d\ln(k)}\bigg)_{k=k_*}   = \frac{dt}{d\ln(Ha)}\frac{d\ln[\Delta_t^2(k_*)]}{dt} \simeq -2\epsilon_E
\,.
\label{nt}
\end{eqnarray}
Next, Eqs.~(\ref{two spectra}) imply that the ratio of the tensor and scalar spectra is,
\begin{equation}
   r(k_*)\equiv \frac{ \Delta_t^2(k_*)}{\Delta_s^2(k_*)} = 16\epsilon_E
\,.
\label{r}
\end{equation}
Finally, the running of the spectral index $n_s$ is,
\begin{equation}
 \alpha(k_*) = \bigg[\frac{d(n_s)}{d\ln(k)}\bigg]_{k=k_*} = -(2\epsilon_E+\xi_E)\eta_E
\,.
\label{running alpha}
\end{equation}
This completes the calculation of the quantities required for slow roll analysis, which is used in the remainder of the paper.
In our plots we shall sometimes express our quantities in terms of the Jordan frame field $\phi=\phi_J$, and sometimes
in terms of the number of e-folds in the Einstein frame, $N_E$. For the latter it is useful to know how to calculate
the field value $\phi_{Je}$ at the end of inflation, which is by convention defined as the field value at which $\epsilon_E=1$.
A cursory look at Eq.~(\ref{eps E})   reveals that $\epsilon_E=1$ when $F={F^\prime}^2/2$, which is equivalent to
the zeros of the following cubic equation for $\phi_J^2$,
\begin{equation}
 8\xi_4^2\phi_J^6 +(8\xi_2\!+\!1)\xi_4M_{\rm P}^2\phi_J^4+(2\xi_2\!+\!1)\xi_2M_{\rm P}^4\phi_J^2-M_{\rm P}^6=0
\,.
\label{phiJe}
\end{equation}
Two of the zeros of this equation are complex and hence unphysical
and one zero is real and positive, representing hence the unique
physical solution defining  the end of inflation. In the following
analyzes we use that solution to signify the end of inflation.

 An important question that needs to be addressed is the validity of slow roll
 approximation. When inflation lasts much longer than $N\simeq 60$,
 it is to be expected that the field will be extremely close to the
 attractor regime described by Eq.~(\ref{Hamilton-Jacobi-type}).
 Under that assumption Eq.~(\ref{Hamilton-Jacobi-type}) can be used
 to test slow roll approximation.
 To get some insight into that question, we shall now study early time evolution of the field, which is described by
Eq.~(\ref{Hamilton-Jacobi-type}). Inserting the {\it Ansatz}, $H_E^2 = H_0^2(1+\zeta \phi_E^2/M_{\rm P}^2)$
into~(\ref{Hamilton-Jacobi-type}) yields a quadratic equation for $\zeta$,
\begin{equation}
  \zeta^2-\frac32\zeta+3\xi_2=0
\,,
\label{exact early evolution}
\end{equation}
where we made use of $V_E(\phi_E)\simeq 3H_0^2[M_{\rm P}^2+2\xi_2\phi_E^2]$, see Eq.~(\ref{VE small field}).
The two solutions are,
\begin{equation}
\zeta_\pm=\frac34\bigg(1\pm\sqrt{1-\frac{16}{3}\xi_2}\bigg) \,.
\label{zeta + -}
\end{equation}
The physically relevant solution is the negative one,
$\zeta=\zeta_-$, as $H_E^2(\phi_E)$ must decrease as $\phi_E^2$
increases.  When $|\xi_2|\ll 1$, $\zeta = 2\xi_2 +{\cal
O}(\xi_2^2)$, so the leading order result in $\xi_2$ reproduces the
slow roll result, and the higher order powers in $\xi_2$ are
corrections to slow roll. Thus, as long as $|\xi_2|\ll 1$, the slow
roll results should be trustable. This is, of course true, provided
the attractor behavior (discussed above) is  realised and
$H_E=H_E(\phi_E)$ does not depend on $\dot\phi_E$. At late times,
when $\phi_E^2\gg M_{\rm P}^2$, the Einstein frame effective
potential reduces to~(\ref{VE large field}). It is well known that
solutions to the Friedmann equations in such an exponential
potential exhibit an attractor
behavior~\cite{Ratra:1987rm,Joyce:1997fc,Glavan:2015ora} in which,
while $\xi_2$ dominates the dynamics,
$\epsilon_E=-8\xi_2/(1-6\xi_2)$, and asymptotically (when $\xi_4$
dominates), $\epsilon_E=4/3$. The slow roll approximation again
reproduces the leading order results: at intermediate times,
 $\epsilon_E=-8\xi_2$, and at late times,  $\epsilon_E=4/3$, leading to an identical conclusion: as long as $|\xi_2|\ll 1$,
slow roll approximation yields approximately correct results. With this in mind, we are ready to proceed to analyze
our inflationary model in slow roll approximation. The analysis that goes beyond slow roll we leave for future work.


%
\begin{figure}[hht]
\begin{minipage}[t]{.45\textwidth}
        \begin{center}
\includegraphics[scale=0.58]{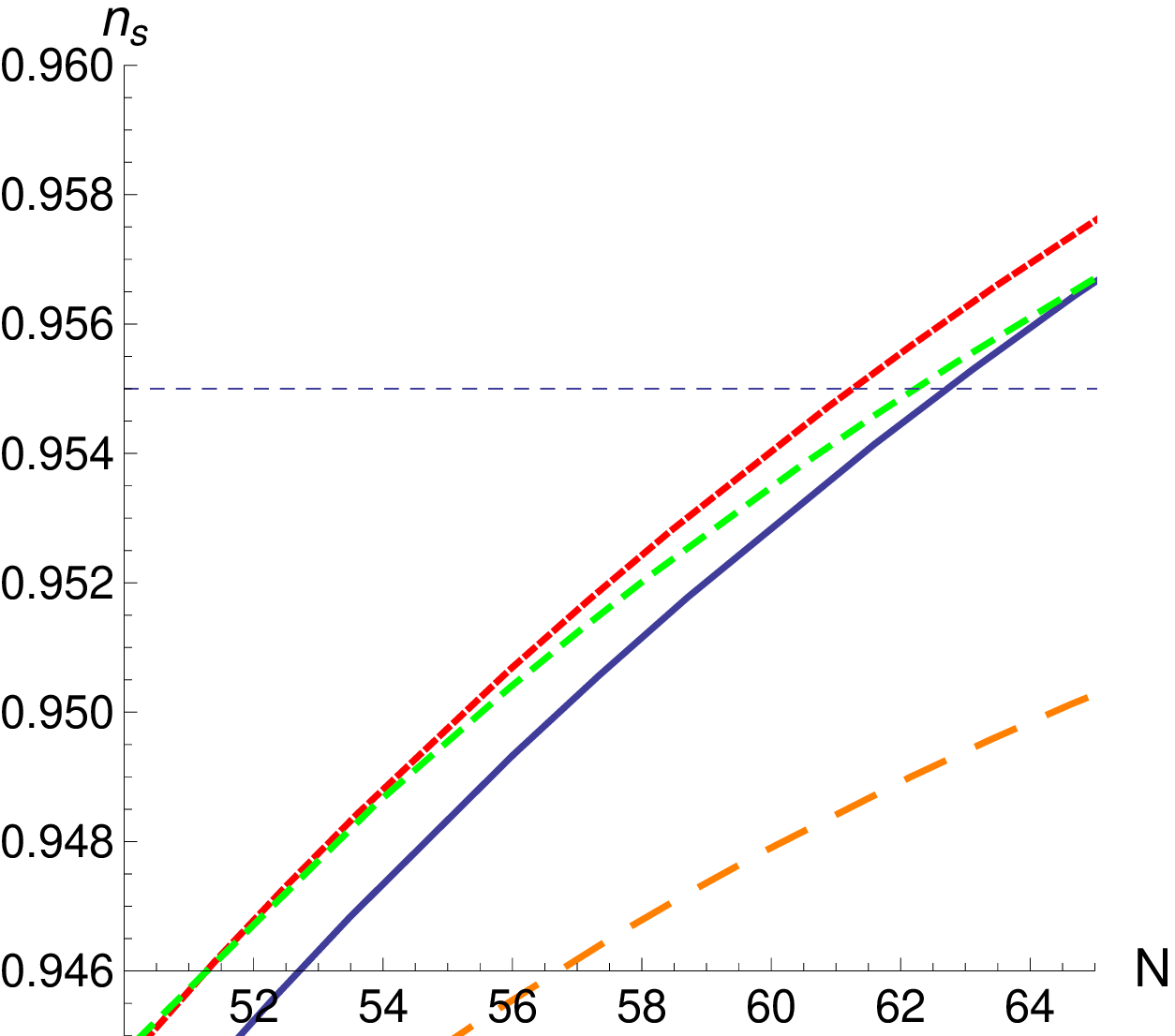}
\end{center}
    \end{minipage}
\begin{minipage}[t]{.45\textwidth}
        \begin{center}
\includegraphics[scale=0.58]{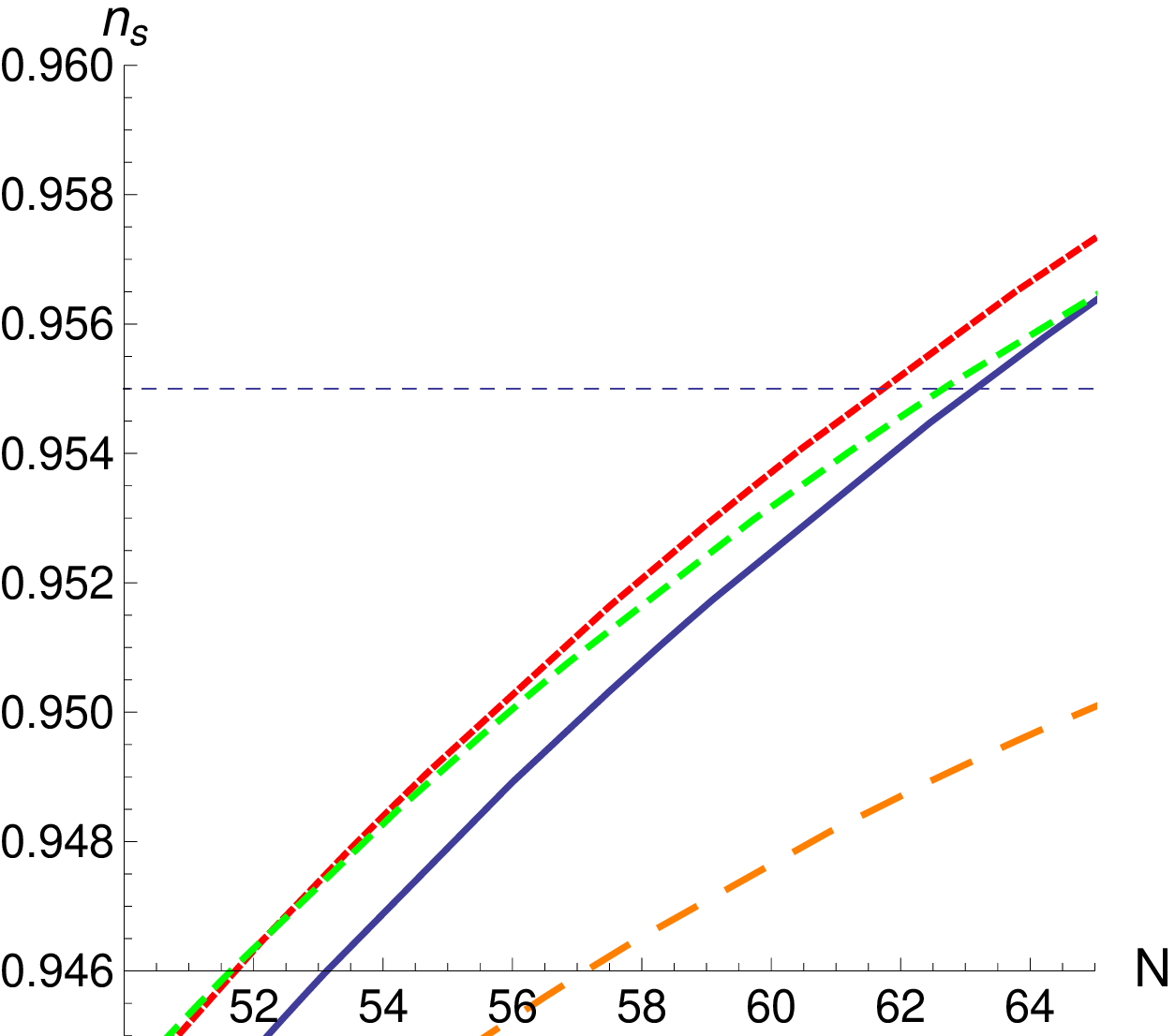}
\end{center}
    \end{minipage}
\caption{Spectral index $n_s$ as a function of the number of e-folds
$N$ (from bottom to top) for $\xi_2=-0.005$ (long orange dashes),
$\xi_2=-0.001$ (blue solid), $\xi_2=-0.003$ (green dashes) and
$\xi_2=-0.002$ (short red dashes). {\it Left panel:} $\xi_4=-0.1$.
{\it Right panel:}  $\xi_4=-0.01$. Numerical investigations show
that the maximum value of $n_s$ is very weakly dependent on $\xi_4$,
and peaks for $\xi_4 \in( -10^{-2},-10^{-1})$. The dependence on
$\xi_2$ is much more pronounced, and
 $n_s$ peaks around $\xi_2\simeq -0.002$  (short red dashes on both left and right panels).}
\label{ns as function of N xi2 and xi4}
\end{figure}

\section{Results}
\label{sec:Results}

In this section we present the principal results for the most
important inflationary observables, which include: the amplitude of
the scalar spectrum $\Delta_s^2$, scalar spectral index $n_s$ and
its (logarithmic) running $\alpha$ and the ratio of tensor and
scalar perturbations, $r= \Delta_t^2/\Delta_s^2$. We do not discuss
separately the tensor spectral index $n_t$ and its running, but
observe in passing that
 (within our approximations) the latter satisfies a consistency relation, $n_t=-2\epsilon_E =-r/8$ and thus, up to a constant rescaling,
$n_t$ is captured by the analysis of $r$.

Figure~\ref{ns as function of N xi2 and xi4} shows the dependence of
the spectral index $n_s$ on the number of e-folds $N$, taking
$\xi_2$ and $\xi_4$ as parameters. The figure shows that $n_s$ peaks
for $\xi_2\simeq -0.002$ (short red dashes), and it is very weakly
dependent on $\xi_4$ (on the left panel $\xi_4=-0.1$ and on the
right panel $\xi_4=-0.01$). Decreasing $|\xi_4|$ further would lead
to smaller values of $n_s$. Since the peak value of $n_s$ in our
model is smaller (by about one standard deviation (dashed blue
horizontal line)) than the central value of $n_s$ obtained by the
Planck collaboration~\cite{Ade:2015lrj}, we conclude that our model
gives the best results when preheating is instant and when
$\xi_2\sim -0.002$ and $\xi_4\sim -0.1$.

\begin{figure}[hht]
\begin{minipage}[t]{.47\textwidth}
        \begin{center}
\includegraphics[scale=0.6]{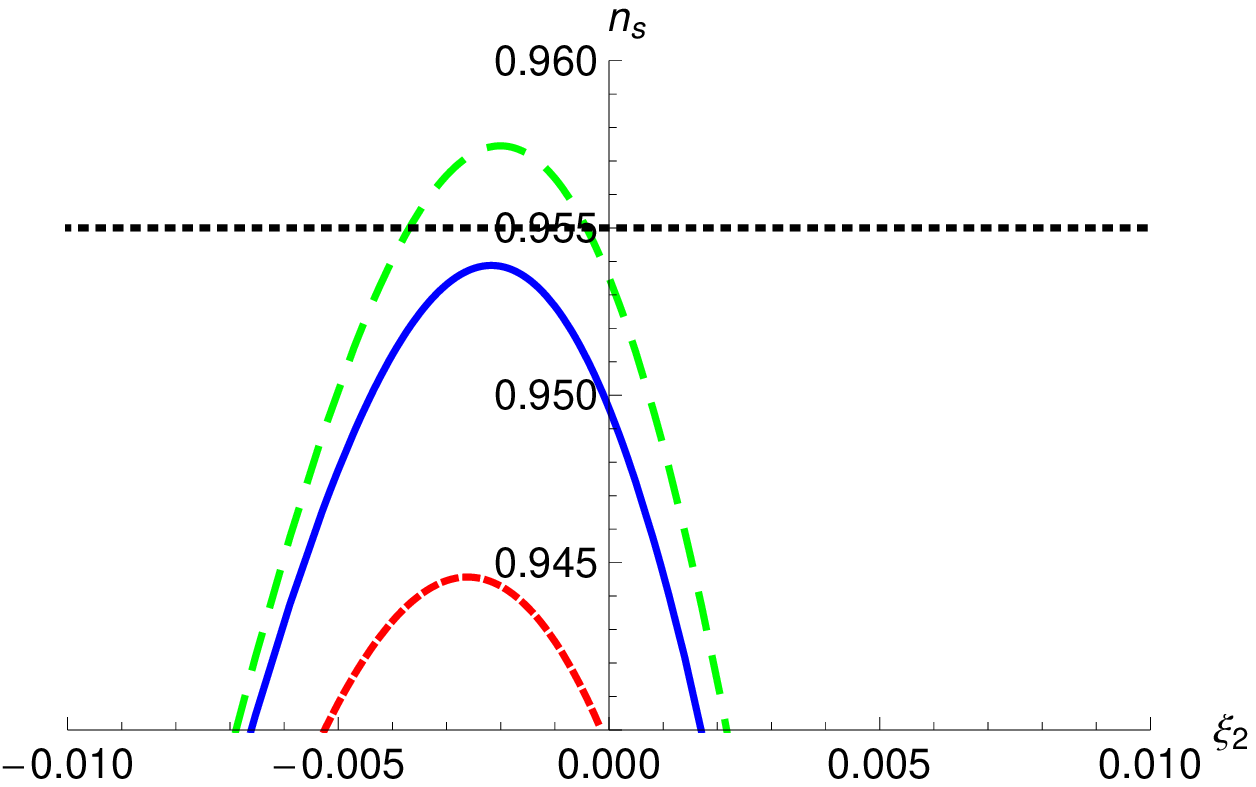}
\end{center}
    \end{minipage}
\begin{minipage}[t]{.45\textwidth}
        \begin{center}
\includegraphics[scale=0.58]{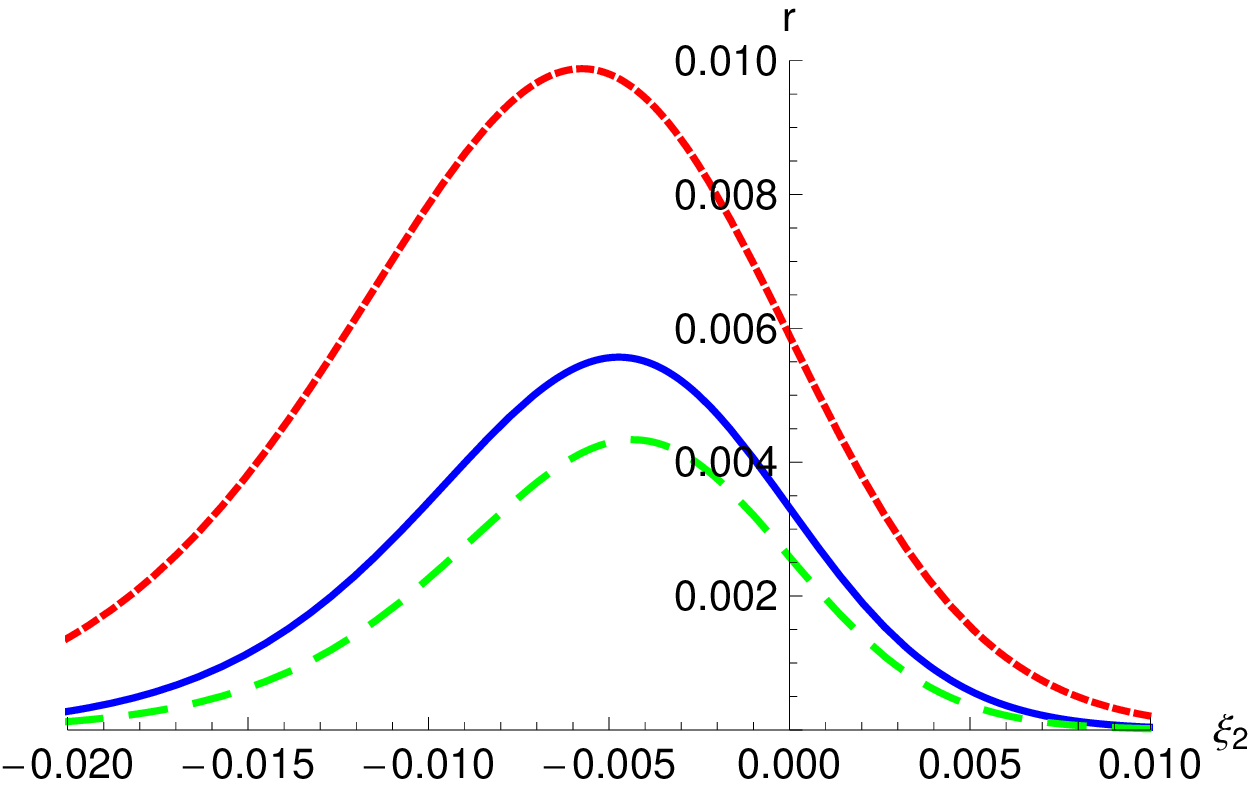}
\end{center}
    \end{minipage}
\caption{{\it Left panel:} The spectral index $n_s$ as a function of $\xi_2$ for
$N=50$ (short red dashes), $N=60$ (blue solid) and $N=65$ (long green dashed).
The maximum value of $n_s$ is very weakly dependent on
$\xi_4$, and peaks for $\xi_4 \in( -10^{-2},-10^{-1})$. The dependence on $\xi_2$ is much more pronounced, and
 $n_s$ peaks around $\xi_2\simeq -0.002$:  $n_s=0.955$ (the value which is about $1\sigma$
lower than the Planck satellite best fit value) when $N\simeq 62$ at
the pivotal scale $k_*=0.05~{\rm Mpc}$. In this graph $\xi_4=-0.02$.
{\it Right panel:} The ratio of the spectra of tensor and scalar
cosmological perturbations $r$ as a function of $\xi_2$ for $N=50$
(short red dashes), $N=60$  (blue solid) and $N=65$  (long green
dashes). The maximum value of $r$ is attained for $\xi_2\sim
-0.005$, and it is approximately inversely proportional to
$|\xi_4|$. This means that, to get values that are observable by the
near future experiments, $|\xi_4|$ needs to be sufficiently small.
Roughly, we have (when $N=60$) $r\sim  10^{-6}/|\xi_4|$, such that
in order to get an observable $r$ one needs $|\xi_4|\leq 10^{-3}$.
In this graph $\xi_4=-0.0002$.} \label{ns and r vs xi2}
\end{figure}
In figure~\ref{ns and r vs xi2} (left panel) we show the dependence
of $n_s$ and $r$ on $\xi_2$ with the number of e-folds $N$ as a
parameter (from bottom up: the curves corresponding to $N=50$, $60$
and $65$ are shown). Figure~\ref{ns and r vs xi2} shows that the
optimal value of $\xi_2$ is about $-2\times 10^{-3}$, which is the
value at which $n_s$ peaks. The peak value of $n_s$ is very weakly
dependent on $\xi_4$ and decreases slowly as $|\xi_4|$ decreases.
The right panel of figure~\ref{ns and r vs xi2} shows the ratio of
the spectral amplitudes of tensor and scalar cosmological
perturbations $r$ as a function of $\xi_2$: $r$ is typically small
and peaks for $\xi_2\simeq -0.005$. Contrary to $n_s$, $r$ shows a
strong (approximately inversely proportional) dependence on $\xi_4$,
such that one can get $r$ as large as $10^{-2}$ when $|\xi_4|\sim
10^{-4}$.

Also from figure~\ref{ns and r vs xi2} one sees that the dependence
of $n_s$ and $r$ on the number of e-folds $N$ (for a sufficiently
large $N$) is approximately,

\beq n_s-1\sim -\frac{p_{n_s}(\xi_2,\xi_4)}{N}\qquad
\mbox{and}\qquad r\sim \frac{p_{r}(\xi_2)/|\xi_4|}{N^3}\, \label{ns
r attractor}
\eeq
where $p_{n_s}$ is weakly dependent on $\xi_2$ and $\xi_4$, and for
the typical choice of the nonminimal couplings taken in this paper,
$p_{n_s}\simeq 2.5$. Likewise, $p_r$ is weakly dependent on $\xi_2$
and $p_r\sim 4$.

\begin{figure}[hht]
\begin{minipage}[t]{.45\textwidth}
        \begin{center}
\includegraphics[scale=0.58]{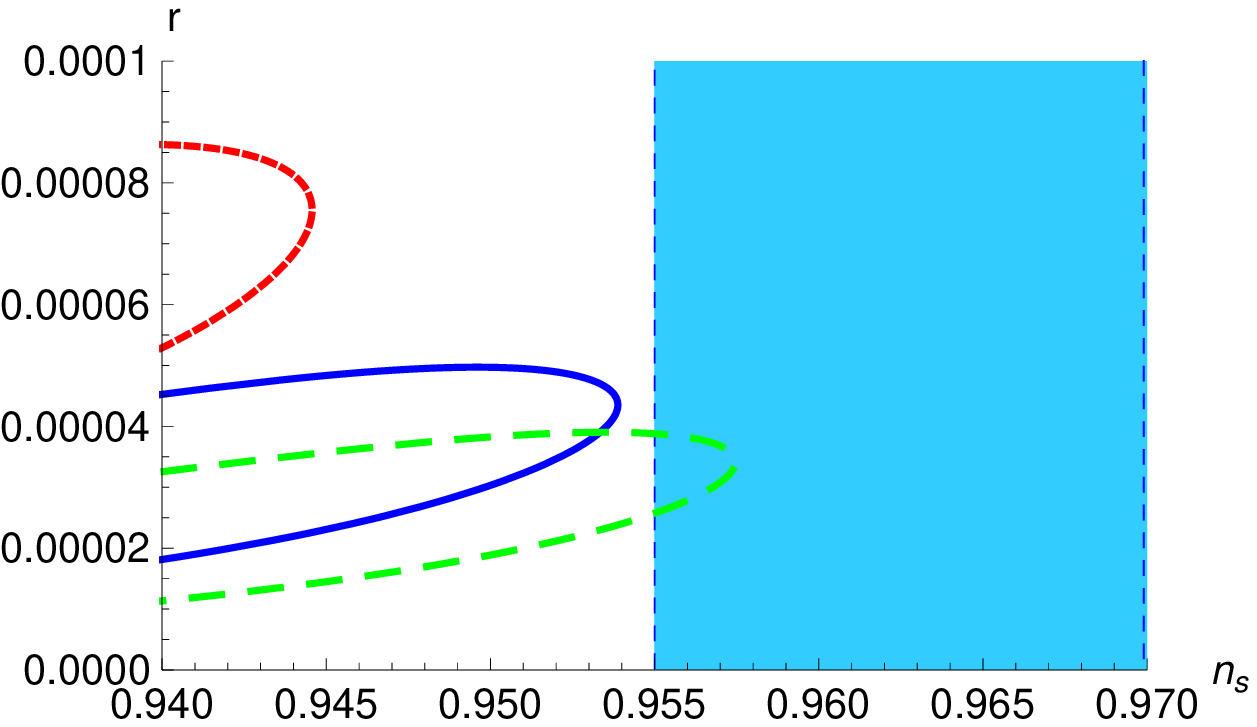}
\end{center}
    \end{minipage}
\begin{minipage}[t]{.45\textwidth}
        \begin{center}
\includegraphics[scale=0.8]{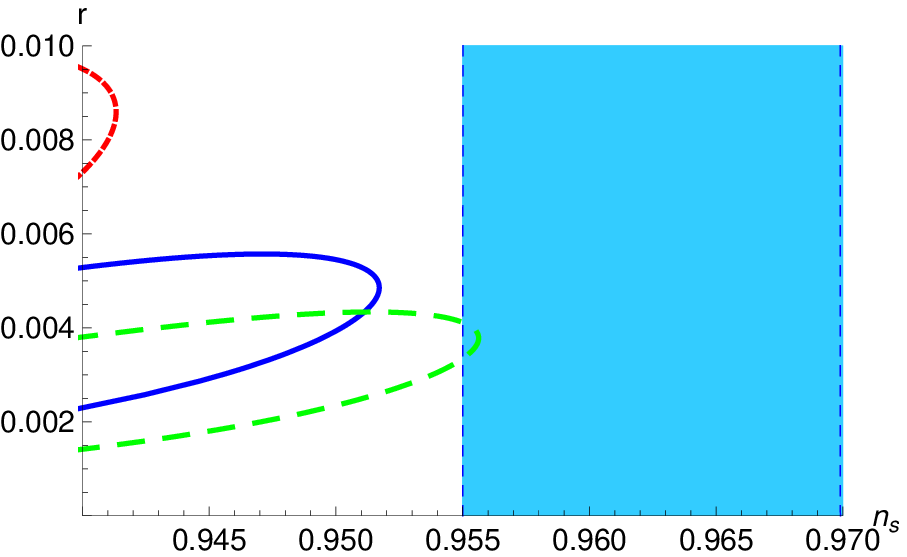}
\end{center}
    \end{minipage}
\caption{The ratio of the spectra of tensor and scalar cosmological
perturbations $r$ {\it versus} spectral index $n_s$ for the number
of e-folds $N=50$  (short red dashes), $N=60$ (blue solid) and
$N=65$  (long green dashes). {\it Left panel:} $\xi_4=-0.02$. {\it
Right panel:}  $\xi_4=-0.0002$. From the figure we see that the
maximum value of $n_s$ grows as $N$ increases, favoring thus models
with a large number of e-folds, such as models with instant
preheating and models which have a postinflationary period of
kination. We also see that $r$ decreases as $N$ increases and as
$\xi_4$ increases. However, making $|\xi_4|$ smaller has as a
consequence a slight reduction of $n_s$.} \label{r vs ns}
\end{figure}
In figure~\ref{r vs ns} we show the ratio of the spectra of tensor
and scalar cosmological perturbations $r$ as a function of the
spectral index $n_s$ for the number of e-folds, $N=50$  (short red
dashes), $60$ (blue solid) and $65$ (long green dashes). We see that
the maximum value of $n_s$ increases as the number of e-folds
increases, and it touches the lower $1\sigma$ observed bound on
$n_s$ (taken from figure~4 of Ref.~\cite{Ade:2015lrj}, from where we
took the $1\sigma$ contours obtained at the optimal value of the
running spectral index) when $N\simeq 62$. The question whether this
high value of $N$ can be obtained within the standard cosmology
(inflation followed by radiation and matter era) is discussed in the
paragraph below. The model favors small values of $r$. An $r$ that
is large enough ($r\sim 10^{-3}-10^{-2}$)
 to be observable by the near future planned missions (such as CORE+ and PRISM~\cite{Andre:2013nfa})
can be obtained at the price of slightly decreasing $n_s$, thus
moving it further away (to about $1.5\sigma$) from the sweet spot,
$n_s\simeq 0.965$.

In conclusion, our analysis shows that, even though our model is
slightly (at $1\sigma$) disfavored by the current data, it is a
viable model of inflation.\\

  A simple calculation shows that the number of e-folds during inflation
that corresponds to some pivotal scale $k_*$ is (see {\it e.g.}
Appendix~B in Ref.~\cite{Glavan:2014uga}),
\begin{equation}
  N_I = \frac{1}{2-\bar\epsilon_I}\Bigg\{\ln\bigg[\frac{H_*}{H_0}\bigg(\frac{k_0}{k_*}\bigg)^\frac{1}{1-\epsilon_I}\bigg]-\frac12 N_m\Bigg\}
\,,
\label{N during inflation}
\end{equation}
where instant preheating is assumed. More accurately: an instant
transition from inflation (during which $\epsilon_E=\epsilon_I$) to
radiation (during which $\epsilon_E=2$) is assumed. In the above
formula $H_0\simeq 68~{\rm km/Mpc/s}$ is the Hubble rate today,
$H_*\simeq 3.4\times 10^{13}~{\rm GeV/\hbar}\simeq 1.6\times
10^{60}~{\rm m/Mpc/s}$ is the Hubble rate at the time when the
pivotal comoving momentum $k_*=0.05~{\rm Mpc^{-1}}$ exits the Hubble
radius during inflation, $k_0=0.00026~{\rm Mpc^{-1}}$ is the
comoving momentum corresponding to the Hubble scale today, and
$N_m\simeq 8.1$ is the number of e-folds during matter era. Once
$N_I$ is known, the number of e-folds during radiation era is easily
calculated from $N_r = (1 -\bar\epsilon_I)N_I - \frac12 N_m$. Taking
$\epsilon_I=\bar\epsilon_I\simeq 0.02$ during inflation gives
$N_I\simeq 59.4$ -- this result is correct provided
$\epsilon_E=\epsilon_I$ stays constant during inflation and then
relatively suddenly (within one or at most a few e-folds) changes at
the end of inflation to $\epsilon_E=2$. More realistically,
$\epsilon_E$ changes gradually during inflation. Indeed, typical
inflationary models predict $\epsilon_E\sim q/N$, where $q$ is a
constant of the order of unity. In these models  $\epsilon_I$ needs
to be replaced by its average value, $\bar\epsilon_I\simeq 0.1$. In
this case Eq.~(\ref{N during inflation}) gives, $N_I\simeq 62.2$.
This is the maximum number of e-folds one can attain during
inflation that correspond to the pivotal momentum $k_*=0.05~{\rm
Mpc^{-1}}$ in standard cosmology.

However, there are non-standard
cosmologies~\cite{Joyce:1997fc,Joyce:2000ag,Spokoiny:1993kt} which
include a period of kination (during which the kinetic energy of a
scalar field dominates the energy density such that $\rho\propto
1/a^6$ and $\epsilon_E\simeq\epsilon_k= 3$). During kination
comoving modes approach the Hubble scale faster than during
radiation or matter era, increasing thus the number of required
inflationary e-folds. For example, when the number of e-folds of
kination is $20\%$ of that in radiation, the number of e-folds
(corresponding to $k_*=0.05~{\rm Mpc^{-1}}$) increases from
$N_I\simeq 62.2$ to $N_I\simeq 66.4$.

In conclusion, a careful calculation shows that the number of
inflationary e-folds corresponding to the pivotal scale
$k_*=0.05~{\rm Mpc^{-1}}$ used by the Planck collaboration is at
most $N_I\simeq 62$ (for standard cosmology), while in nonstandard
cosmologies (with {\it e.g.} a period of kination) it can be larger.
For these reasons in our figures we show results not just for $N=50$
and $N=60$, but also for $N=65$.\\

 Let us now try to figure out what the current data can tell us about the magnitude of the cosmological constant $\Lambda$
in our inflationary model. Recall that we know~\cite{Ade:2015lrj}
that the amplitude of the scalar power spectrum (the COBE
normalization) at the pivotal scale $k_*=0.05~{\rm Mpc^{-1}}$ is
$\Delta_s^2(k_*)=(2.20\pm 0.09)\times 10^{-9}$. On the other hand,
combining Eqs.~(\ref{EOM: Einstein frame:2}), (\ref{two spectra})
and~(\ref{r}) gives,
\begin{equation}
  \frac{\Lambda}{M_{\rm P}^2} = \frac{F^2}{M_{\rm P}^4}\frac{3\pi^2}{2}\Delta_{s}^2(k_*)r
    \simeq \frac{3\pi^2}{2}\Delta_{s}^2(k_*)r =(3.25\pm 0.13)\times 10^{-8}\,r
\,,\quad r \sim \frac{10^{-6}}{|\xi_4|}
\,,
\label{cosmological constant value}
\end{equation}
where in the second equality we used the approximation, $F\simeq M_{\rm P}^2$.
To investigate whether the value of this cosmological constant is at the grand unified scale (GUT),
let us define the GUT energy density as, $\rho_{\rm GUT}\equiv E_{\rm GUT}^4 = \Lambda M_{\rm P}^2$,
from which one gets,
\begin{equation}
  \frac{\Lambda}{M_{\rm P}^4} \simeq 2.84\times 10^{-10}\bigg(\frac{E_{\rm GUT}}{10^{16}~{\rm GeV}}\bigg)^4
\,.
\label{cosmological constant value:2}
\end{equation}
 Comparing this with~(\ref{cosmological constant value}) gives the following estimate of the grand unified scale producing $\Lambda$,
\begin{equation}
  \frac{E_{\rm GUT}}{10^{16}~{\rm GeV}}\simeq 3.27\times r^{1/4}
\label{cosmological constant value:2}
\end{equation}
which yields $E_{\rm GUT}\simeq 10^{16}~{\rm GeV}$ when $r\sim 10^{-2}$. Thus it is fair to say that for a rather broad range of $r$'s
the cosmological constant in our model is at the grand unified scale.

\begin{figure}[hht]
\begin{minipage}[t]{.45\textwidth}
        \begin{center}
\includegraphics[scale=0.5]{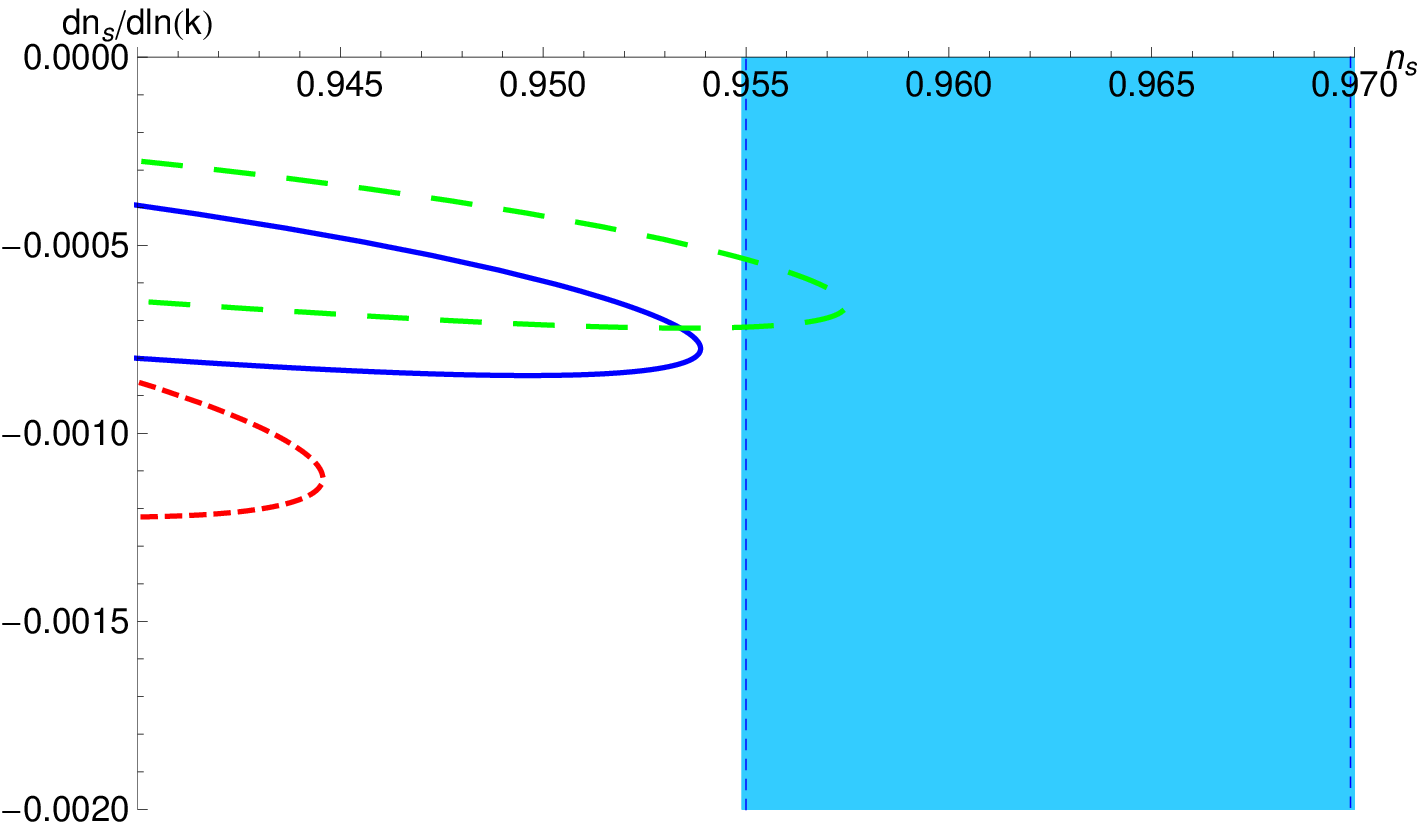}
\end{center}
    \end{minipage}
\begin{minipage}[t]{.45\textwidth}
        \begin{center}
\includegraphics[scale=0.5]{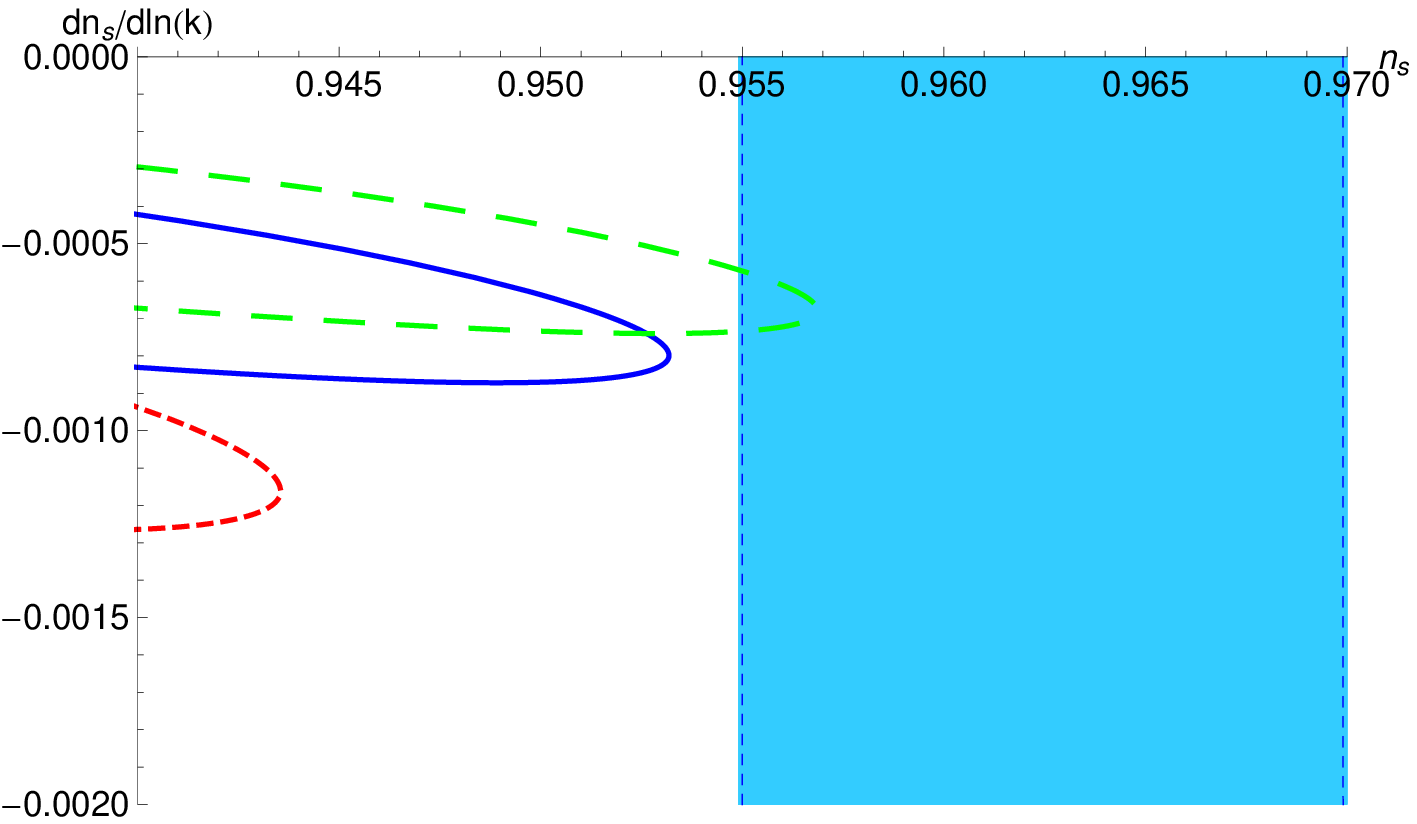}
\end{center}
    \end{minipage}
\caption{{\it Left panel:} The running of the spectral index $\alpha=dn_s/d\ln(k)$ as a function of $n_s$ for
$N=50$ (short red dashes), $N=60$ (blue solid) and $N=65$ (long green dashed).
Both the maximum value of $n_s$ and $\alpha$ are very weakly dependent on
$\xi_4$. The typical value of the running is about $\alpha\sim -10^{-3}$, which is consistent with the current
Planck data and it is about a factor of a few smaller in value than the central value favored by
the Planck (and other) data, $\alpha\sim -0.003\pm 0.007$. In this graph $\xi_4=-0.02$.
{\it Right panel:} The same as in the left panel but with $\xi_4=-0.2$.
The value of the running $|\alpha|$ grows slowly as $|\xi_4|$ increases.}
\label{alpha vs ns}
\end{figure}
 Figure~\ref{alpha vs ns} shows how the running of the spectral index  $\alpha=dn_s/d\ln(k)$
depends on the spectral index $n_s$. While the dependence of $n_s$
on $\xi_2$ and $\xi_4$ is by now familiar, we see from the figure
that $\alpha$ depends very weakly on $\xi_4$ (increasing slowly as
$|\xi_4|$ increases), and for $N\simeq 60$ peaks at a value,
$\alpha\sim -0.0008$, which is to be compared with the value
observed by the Planck collaboration, $\alpha=-0.003\pm
0.007$~\cite{Ade:2015lrj}. Therefore, the spectral index running in
our model is consistent with the current data and it is potentially
observable provided the error bars decrease by about a factor of 10.
It is unlikely that such an accuracy in $\alpha$ can be attained by
the near future CMB missions. Therefore, observing a running
different from zero in the near future would be tantamount to ruling
out our model.

\begin{figure}[hht]
\begin{minipage}[t]{.47\textwidth}
        \begin{center}
\includegraphics[scale=0.58]{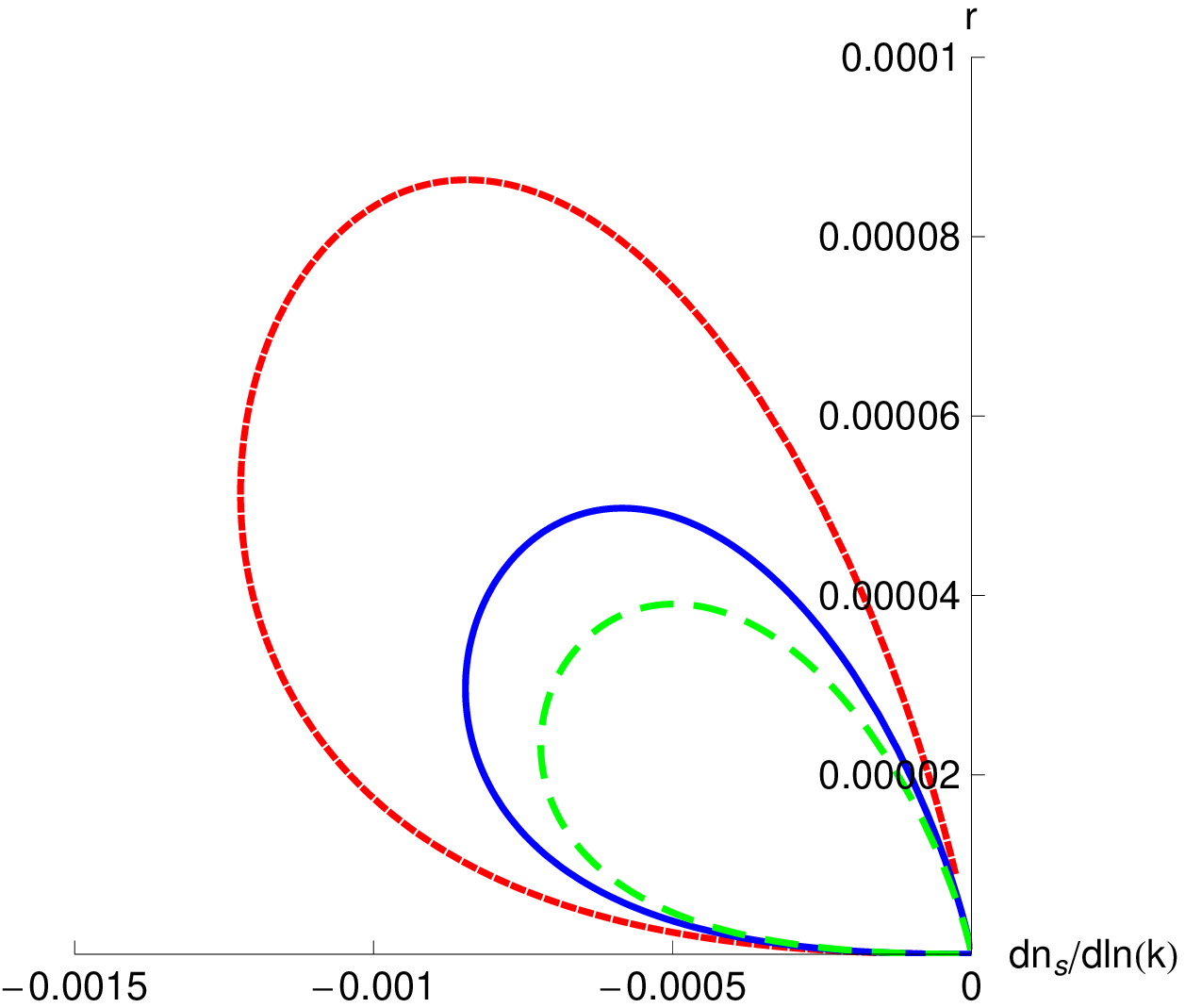}
\end{center}
    \end{minipage}
\begin{minipage}[t]{.47\textwidth}
        \begin{center}
\includegraphics[scale=0.58]{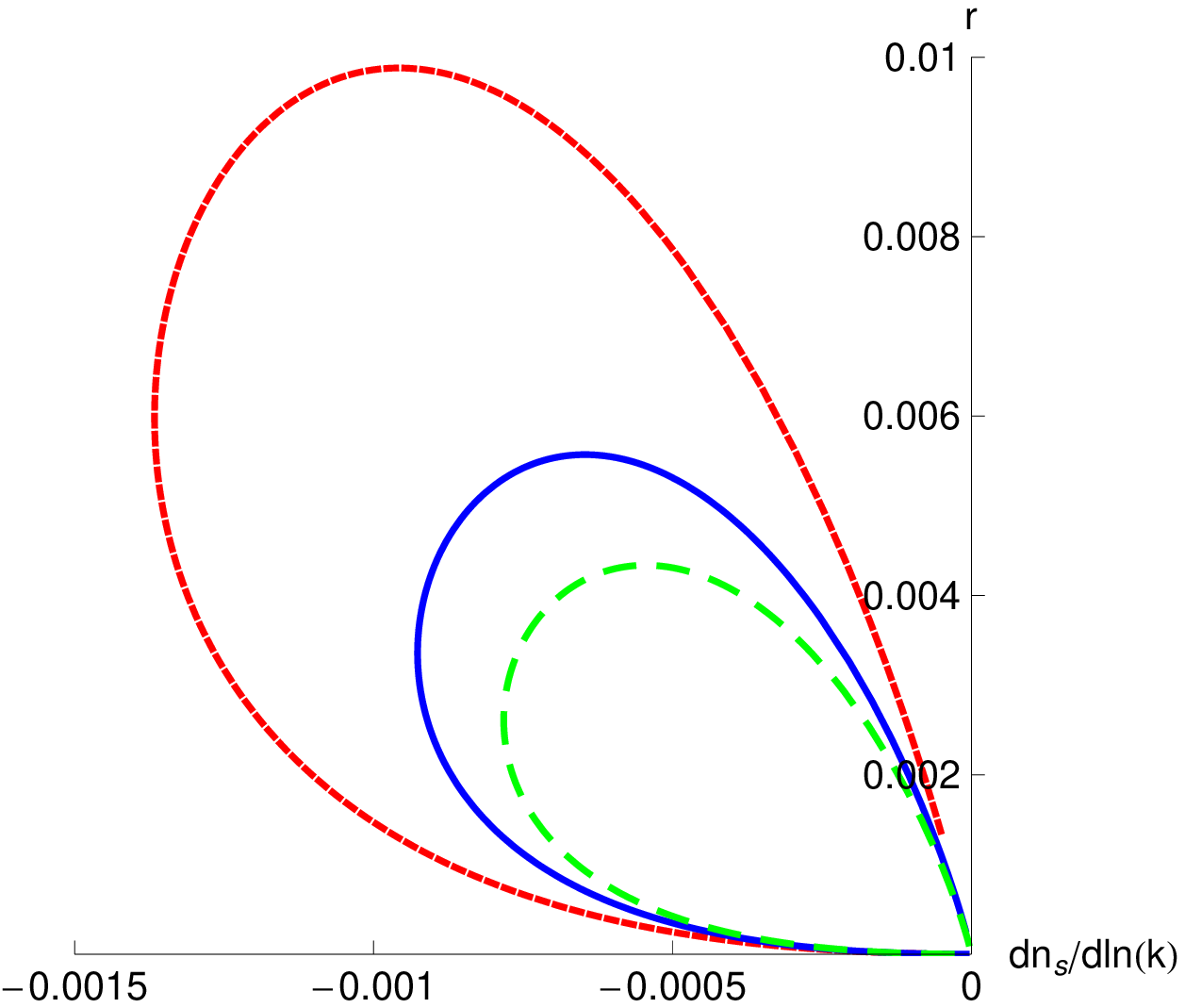}
\end{center}
    \end{minipage}
\caption{{\it Left panel:} The tensor-to-scalar ratio $r$ {\it vs}
the running of the spectral index $\alpha=dn_s/d\ln(k)$ for $N=50$
(short red dashes), $N=60$ (blue solid) and $N=65$ (long green
dashed). For these curves $\xi_4=-0.02$. {\it Right panel:} The same
as in the left panel but with $\xi_4=-0.0002$. Note that as $N$
increases the values of $r$ and $\alpha$ decrease. A comparison of
the left and right panels reveals that $r\propto 1/|\xi_4|$, which
was already pointed out above. For wide ranges of values of $\xi_2$
and $\xi_4$ the values of $\alpha$ and $r$ are small enough to be
consistent with the current observations.} \label{r vs alpha}
\end{figure}
In figure~\ref{r vs alpha} we show the dependence of $r$ on $\alpha=dn_s/d\ln(k)$ with $N$ as a parameter
($N=50$ (short red dashes), $N=60$ (blue solid) and $N=65$ (long green dashed)). On the left panel $\xi_4=-0.02$, while on the right
panel  $\xi_4=-0.0002$. The figure shows that, while the running $\alpha$ very weakly dependents on $\xi_4$,
$r$ is approximately inversely proportional to $\xi_4$. If future observations show that $r\sim 10^{-2}$ that would mean
that $\xi_4$  would have to be small ({\it e.g.} $\xi_4\sim -10^{-4}$)  and that $n_s$ would have to be below about $0.950$.


 \section{Discussion}
 \label{sec:Discussion}

 In this paper we analyze a novel inflationary model, where inflation is driven by a (Jordan frame) cosmological constant
and a nonminimally coupled scalar field plays the role of the
inflaton. The model is inspired by the recent
work~\cite{Glavan:2015ora}, where it was argued that, when viewed in
the context of a nonminimally coupled scalar, a Jordan frame
cosmological constant can be dynamically relaxed to zero (from the
point of view of the Einstein frame observer). The model is analyzed
in slow roll approximation, whose accuracy is (to a certain extent)
tested. Our analysis shows that we can get the spectral index
consistent with current observations, albeit the maximum value of
the spectral index is about one standard deviation below the
observed value, see figure~\ref{ns as function of N xi2 and xi4}.
The value of the tensor-to-scalar ratio $r$ is typically small, see
figure~\ref{ns and r vs xi2}. Since $r$ is inversely proportional to
the quartic nonminimal coupling $|\xi_4|$, it can be enlarged by
decreasing the value of $|\xi_4|$ to obtain an $r$ that is
observable by the planned CMB experiments, but the price to pay is a
smaller $n_s$. The running of the spectral index $\alpha$ is
negative, see figures~\ref{alpha vs ns} and~\ref{r vs alpha}, but
the typical amplitude of the running is by about one order of
magnitude below the sensitivity of the current CMB data.

 It is worth noting that the value of the cosmological constant is to a large extent determined by the COBE
normalization and it is of the order of the GUT scale, {\it i.e.}
$\Lambda/(8\pi G_N)\sim E_{\rm GUT}^4\sim 10^{16}~{\rm GeV^4}$, and
hence it can be nicely attributed to the value attained at a GUT
transition (both from the Higgs potential as well as from the
contributions  generated by the particle masses). A second nice
feature of the model at hand is that the model works for a large
class of initial conditions. Namely, inflation naturally begins from
a chaotic state, in which the total (averaged) energy in
fluctuations scales as $\langle\rho\rangle\propto 1/a^4$ and during
which the average field value is naturally small, $\langle
\phi\rangle\ll M_{\rm P}$ (this is so because the nonminimal
coupling plays no role as the Ricci scalar is small, $\langle
R\rangle\sim 0$). Sometime after the GUT transition the cosmological
constant starts dominating the energy density, and one enters (slow
roll) inflation. Inflation is terminated as $\epsilon_E\sim 1$; at
asymptotically late times $\epsilon_E\simeq 4/3$. Graceful exit and
preheating is solved by suitably coupling the scalar field to
matter, for details see Ref.~\cite{Glavan:2015ora}. Another
advantage of our model is in that there is no need to fine tune the
potential to zero at the end of inflation, thus getting rid of one
of the major fine tuning problems of scalar inflationary models.

 From our analysis it is not completely clear how accurate is the slow roll approximation utilized in this paper.
For that reason we are working on studying predictions of the inflationary model presented here by
using exact solutions of the Friedmann equations~(\ref{EOM: Einstein frame:1}--\ref{EOM: Einstein frame:3}).
One hope is that, performing an exact analysis will allow us to obtain values for $n_s$ and $r$ that are
closer to the (central) values favored by current observations, and thus get an even better agreement with
the data.

\subsection*{Acknowledgements}
This work is part of the D-ITP consortium, a program of the
Netherlands Organization for Scientific Research (NWO) that is
funded by the Dutch Ministry of Education, Culture and Science
(OCW). A.M. is funded by NEWFELPRO,  an International Fellowship
Mobility Programme for Experienced Researchers in Croatia and by the
D-ITP.

\end{document}